\documentstyle[12pt]{article}
\begin{document}
\title{Finite Mass Isothermal Spheres and the Structure of Globular
Clusters}
\author{Jes Madsen\\ Institute of Physics and Astronomy\\ University of
Aarhus\\ DK-8000 Aarhus C\\ Denmark\\ e-mail: jesm@dfi.aau.dk\\ \\ \\
\\ To appear in: Monthly Notices of the Royal Astronomical Society}
\date{}
\maketitle
\begin{abstract}
Utilizing a recently derived extension
of the Maxwell-Boltzmann distribution to low occupation
numbers (Simons 1994) this investigation
discusses the structure of stellar dynamical isothermal spheres.
The resulting models, which resemble King models, 
constitute a new
sequence of physically well-motivated spherical equilibrium
configurations of finite mass, and give nice fits to pre-core 
collapse globular cluster data.
\vskip 1cm
{\it Subject headings:} Galaxy: Globular Clusters: General; Celestial
Mechanics, Stellar Dynamics; Galaxies: Elliptical and Lenticular, Cd;
Galaxies: Fundamental Parameters.
\end{abstract}
\section{Introduction}
Globular clusters show lots of fine-structure related to
mass-segregation, velocity anisotropies, core collapse, etc., but as a first
approximation their structure (at least prior to core collapse) 
is surprisingly well described by simple
King models based on truncated isothermal spheres (see Elson, Hut \&
Inagaki 1987 for a review).

Globular clusters are small enough for two-body encounters among the
stars to play a significant role in their evolution, so
Maxwell-Boltzmann statistics (isothermal spheres)
might be expected to be relevant for them,
just like in the case of elastically scattering molecules in a gas.
Elliptical galaxies are too large for two-body encounters among the
stars to be important, but if they formed through processes resembling
violent relaxation, Maxwell-Boltzmann statistics could play a role
there as well (see Binney \& Tremaine 1987 for an overview and
references on these topics). 

It has long been recognized, that a self-gravitating stellar dynamical
system described by a Maxwell-Boltzmann distribution of the form
$f(E)\propto \exp (-\beta E)$, where $E$ is the stellar energy, has
infinite mass, and therefore can at most provide an approximate
description of, e.g., globular clusters or elliptical galaxies.

In the context of model-building,
this has been remedied in several ways in terms of
introducing cutoffs in the distribution function. Most notably the King
models (King 1966) based on the lowered isothermal sphere 
with $f_K(E)\propto \left( \exp
(-\beta E) -\exp (-\beta\Phi_0)\right)$, where $\Phi_0$ 
is a constant discussed below. 
King models fit the observational properties of
globular clusters very well, and to less extent elliptical galaxies,
but their somewhat {\it ad hoc\/} nature is slightly disturbing.

Recently Simons (1994)
derived an exact expression for Maxwell-Boltzmann statistics by a method
which correctly accounts for
discreteness effects by allowing macrocells in phase space to be
occupied by integer numbers of particles only. The resulting
distribution, which forms the basis for the stellar dynamical
applications in the present investigation, is given by
\begin{equation}
F_p=\left[ Ag_p\exp (-\beta E_p) \right] ,
\end{equation}
where $[b]$ is the value of $b$ rounded down to the nearest integer,
$F_p$ and $g_p$ are the number of particles and the number of available
states with energy $E_p$, and $A$ and $\beta$ are constants to be
determined from the conservation equations $\sum F_p =N$, $m\sum F_pE_p
=E_{\rm tot}$, where $N$ and $E_{\rm tot}$ denote total number of particles 
and total energy, and $m$ is the particle mass. In the present investigation
all particles are assumed to be of equal mass, and the word ``energy''
is used as a short synonym for energy per unit mass, except $E_{\rm
tot}$, which is a true energy.

It can be shown explicitly (Simons, 1994), that this distribution
maximizes entropy under the implicit assumption of fixed energy levels. 
For self-gravitating systems no maximum entropy state exists, since the
entropy can be increased by concentrating the core and letting some
particles go into a diffuse halo. But in the case of e.g.\ globular
clusters undergoing slow evolution dominated by two-particle
encounters (i.e.\ prior to more violent phenomena like core collapse)
the gravitational potential is almost time-independent, and stars can be
expected to adjust into a distribution close to that which
maximizes the entropy for a fixed potential.
This is the distribution given in equation (1).

In the following sections equation (1) will be discussed in the context of
stellar dynamics, fits of globular cluster data will be obtained, and
the fundamental parameters and assumptions in the model will be
described.

\section{The Poisson equation for Simons-Maxwell-Boltzmann statistics}
To find the structure of a spherically symmetrical stellar system
described by equation (1) one has to solve Poisson's equation on the form
\begin{equation}
{1\over r^2}{d\over{dr}}\left( r^2{{d\Phi}\over{dr}}\right)=4\pi G\rho ,
\end{equation}
where the density is given by $\rho =\int f(E)d^3v$, and
$f$ denotes a mass per length cubed per velocity cubed; for
stars of identical mass such a coarse-grained distribution is obtained 
from equation (1) by multiplying with $\eta /g_p$, where $\eta$ is the phase
space density of a microcell occupied by one star. 
Macrocells will be constructed 
so as to contain the same number of microcells, $g_p$.

It is customary to introduce a relative gravitational potential, $\Psi$,
related to the true potential $\Phi$ via $\Psi \equiv -\Phi +\Phi_0$,
and a relative energy $\epsilon\equiv -E+\Phi_0=\Psi-{1\over 2}v^2$,
with the constant $\Phi_0$ chosen so that $f>0$ for $\epsilon >0$, and $f=0$
otherwise. Then $\Psi$ satisfies Poisson's equation on the form
$\nabla^2\Psi=-4\pi G\rho$ with $\Psi\rightarrow\Phi_0$ for
$r\rightarrow\infty$, or more explicitly
\begin{equation}
{1\over r^2}{d\over{dr}}\left( r^2{{d\Psi}\over{dr}}\right)=
-16\pi^2G\int_0^{\Psi} f(\epsilon )\left( 2(\Psi -\epsilon )
\right)^{1/2}d\epsilon ,
\end{equation}
with
\begin{equation}
f(\epsilon )={\eta\over{g_p}}\left[ \exp (\beta\epsilon)\right] .
\end{equation}
This corresponds to $\Phi_0=\ln (Ag_p)/\beta$.

Introducing a dimensionless radius $s\equiv r/a$, where $a\equiv \left(
\beta^{1/2}g_p/(G\eta)\right)^{1/2}$, and scaling potentials and
energies in units of $\beta$ one finally gets a dimensionless Poisson
equation
\[
{1\over s^2}{d\over{ds}}\left( s^2{{d(\beta\Psi)}\over{ds}}\right)=
-16\pi^2\int_0^{\beta\Psi} [\exp (\beta\epsilon)]
(2(\beta\Psi -\beta\epsilon ))^{1/2}d(\beta\epsilon)\]
\[
=-16\pi^2\left( \sum_{j=1}^{j_{max}-1} \int_{\ln j}^{\ln (j+1)} j
(2(\beta\Psi -\beta\epsilon ))^{1/2} d(\beta\epsilon ) 
+\int_{\ln j_{max}}^{\beta\Psi} j_{max}
(2(\beta\Psi -\beta\epsilon ))^{1/2} d(\beta\epsilon)\right)\]
\begin{equation}
=-{{2^{11/2}\pi^2}\over 3}\sum_{j=1}^{j_{max}} (\beta\Psi -\ln j)^{3/2} .
\end{equation}
Here $j_{max}=[\exp (\beta\Psi)]$.

A solution of the dimensionless Poisson equation exists for each choice
of scaled central potential, $\beta\Psi (0)$, subject to the boundary condition
$\beta\Psi '(0)=0$ (no force at the center).
The dimensionless potential decreases monotonously
and approaches a negative constant, $\beta\Phi_0$, for $s\rightarrow\infty$.
The corresponding density (in units of $\eta/(g_p\beta^{3/2})$) 
decreases from a central value, $\rho^*$, and reaches
0 when $\beta\Psi =0$ at the dimensionless tidal radius, $s_t$. 
This is the major difference relative to the
``traditional'' isothermal sphere, where the density only asymptotically
approaches zero. Thus the total mass, $M=\int 4\pi\rho r^2dr$, is finite.

To illustrate the importance of discreteness effects consider first the
simplified situation where the physical gravitational potential as a
function of radius, $\Phi (r)$, is fixed a priori (with $\Phi
(r)\rightarrow 0$ for $r\rightarrow\infty$). In this case one can 
specify a partitioning of $(\bf{x,v})$ phase space, including
specification of energy levels, $E_p$, with equal
numbers of available states, $g_p$. One may choose the numbering of
levels such that $E_1$ is the lowest level, i.e.\ closest to $\Phi (0)$,
etc. Energy level $p$ is occupied by $F_p$ particles, where 
$F_p=[Ag_p\exp (-\beta E_p)] \equiv [\exp (\beta \epsilon_p)] \equiv
[\exp (\beta\Phi_0)\exp(-\beta E_p)] \equiv [\exp(\beta\Psi(0))\exp
(\beta(\Phi(0)-E_p))]$,
using the identities $Ag_p\equiv\exp (\beta\Phi_0)$, and $\epsilon_p
\equiv \Phi_0-E_p \equiv \Psi(0)+\Phi(0)-E_p$.

The classical isothermal sphere has (fractional) occupation of energy
levels all the way to $E_p\rightarrow 0$ for $p\rightarrow\infty$ (with
particles reaching $r\rightarrow\infty$), so to rescale the
gravitational potential one would normally use
$\Phi_0=0$, $\Psi(r)=-\Phi(r)$. For the
true isothermal spheres of the present investigation there is a finite
number of particles, and levels are not fractionally occupied, so only levels
with $p$ below some finite $p_{\max}$ contribute, corresponding to
$E_p\leq E_{p_{\rm max}}<0$. Choosing $\Phi_0=E_{p_{\rm max}}$ the rescaled
potential $\Psi(r)\equiv\Phi_0-\Phi(r)$ now has the property that it is
positive for exactly that range of radii where particles with
$\Phi(0)\leq E_p\leq E_{p_{\rm max}}$ ($\Psi(0)\geq\epsilon_p\geq 0$) are able
to move.

The total mass $M$ and physical energy (not per unit mass)
$E_{\rm tot}$ are given by
\begin{equation}
M\equiv mN=m\sum_p F_p=m\sum_p\left[\exp(\beta\Psi(0))\exp(\beta
(\Phi(0)-E_p))\right]
\label{masstot}
\end{equation}
\begin{equation}
E_{\rm tot}=m\sum_p E_pF_p=m\sum_p E_p\left[\exp(\beta\Psi(0))\exp(\beta
(\Phi(0)-E_p))\right] .
\end{equation}

If $\beta$ were fixed, then a system with fixed total mass $M$ could be
realized for given particle mass, $m$, with that choice of $\Psi(0)$
which obeys equation (\ref{masstot}). Clearly, lowering $m$ (and thereby
increasing the number of particles, $N$) corresponds to picking a higher
value of $\Psi(0)$ to increase the sum. This is equivalent to choosing
the (negative) value $\Phi_0$ closer to 0, allowing particles to occupy
states with higher ``quantum number'' $p$. Thus a decrease in particle
mass is directly related to an increase in the scaled potential, $\beta\Psi$,
for fixed total mass. As one sees from the figures, solutions with
higher dimensionless central potential $\beta\Psi(0)$ (corresponding as
just argued to more particles for fixed total mass, physical potential and
phase space partitioning) are as expected closer to the standard
isothermal sphere solution. I.e., discreteness effects are less
conspicuous for large $N$.

In reality, $\beta$ would not be fixed in this procedure since the
solution must also obey the equation for $E_{\rm tot}$, but since (for
dimensional reasons) $E_{\rm tot}=M\beta^{-1}h(\beta\Psi(0))$, where $h$
is a slowly varying function of $\beta\Psi(0)$, $\beta$ does not change
much if $M$ and $E_{\rm tot}$ are fixed.

For the self-consistent case where the potential $\Phi(r)$ is given by
Poisson's equation rather than fixed a priori, the picture is more
complicated, but $\Phi(r)$ is not strongly varying with $N$ for fixed $M$
and $E_{\rm tot}$, so qualitatively the argument above remains
valid---discreteness effects are more pronounced for low $N$, and for
fixed $M$ and $E_{\rm tot}$ (and partitioning of phase space) a change
in $m$ ($N$) corresponds to a change in ``shape parameter''
$\beta\Psi(0)$ for the isothermal spheres.

\section{Comparing with observations}
As explained above, the new Maxwell-Boltzmann statistics in a natural
way accounts for finite mass systems. However, many parameters appear
($A$, $g_p$, $\eta$, $\beta$), so it is not obvious that much has been
gained in terms of constructing useful stellar dynamical models. It will
now be demonstrated how these parameters are related to observable
properties of, e.g., globular clusters.

Figure 1 shows the density profiles for several choices of $\beta\Psi (0)$.
Densities are scaled in units of the central density, and radii in units
of the King radius,
\begin{equation}
r_0\equiv\sqrt {{9}\over{4\pi\beta G\rho (0)}} = \sqrt{9\over{4\pi\rho^*}}a .
\end{equation}
Also shown in Figure 1 is the standard isothermal sphere, which the
present models approach for $\beta\Psi (0)\rightarrow\infty$, and the
corresponding King models. The similarities are 
striking---not surprisingly, since King models also have an energy
cutoff at $E=\Phi_0$ and an element of rounding down ($f_K\propto
(\exp (\beta\epsilon )-1)$ rather than $[\exp (\beta\epsilon )]$). 

Equally similar are the RMS velocities at a given spatial radius in
Figure 2.
(For the full Simons-Maxwell-Boltzmann statistics one can
show that 
\begin{equation}
\overline{v^2}(r) ={6\over 5}\beta^{-1}{{\sum_j(\beta\Psi (s)-\ln
j)^{5/2}}\over{\sum_j(\beta\Psi (s)-\ln j)^{3/2}}} .
\end{equation}
The velocities shown in Figure 2 are $\sqrt {\overline{v^2}(r)\beta
/3}$\ ).

Figure 3 shows the so-called concentration parameter defined as
$c\equiv\log_{10}(r_t/r_0)$. The value of $c$ is typically 0.3 units
lower than the corresponding King model with the same $\beta\Psi(0)$.
King models fit globular
cluster data quite well for $0.75<c<1.75$ and elliptical galaxies
moderately well for $c>2.2$ (Binney \& Tremaine 1987; Mihalas \&
Binney 1981). Similar limits are reasonable for the isothermal spheres.
This corresponds to $5\leq\beta\Psi(0)\leq 9$ and
$\beta\Psi(0)\geq 11$ for the isothermal spheres.

Each choice of dimensionless central potential, $\beta\Psi (0)$, 
leads to values for the
dimensionless central density, $\rho^*\equiv\rho (0)g_p\beta^{3/2}/\eta$,
the dimensionless total mass, $M^*\equiv MG\beta /a$, the dimensionless tidal
radius $s_t\equiv r_t/a$, and the dimensionless, central line-of-sight velocity
dispersion, $\sigma^{*2}\equiv\beta\sigma^2=2\int_0^{s_t}\sum_j(\Psi (s)-\ln
j)^{5/2}ds/5\int_0^{s_t}\sum_j(\Psi (s)-\ln j)^{3/2}ds$.

Explicitly expressing observable quantities in terms of $\eta$ (in units
of $M_\odot$pc$^{-3}$(km/s)$^{-3}$), $\beta$ (in units of
(km/s)$^{-2}$), and $g_p$, one gets
\begin{equation}
\rho (0) =\rho^* \eta g_p^{-1}\beta^{-3/2} M_\odot\  \mbox{\rm pc}^{-3},
\end{equation}
\begin{equation}
M=3.546\times 10^3 M^* \beta^{-3/4}\eta^{-1/2}g_p^{1/2}\  M_\odot ,
\end{equation}
\begin{equation}
r_t=15.25s_t\beta^{1/4}\eta^{-1/2}g_p^{1/2}\  \mbox{\rm pc},
\end{equation}
and
\begin{equation}
\sigma^2=\sigma^{*2}\beta^{-1}\  \mbox{\rm (km/s)}^2 .
\end{equation}

Notice that $\eta$ and $g_p$ always occur in the combination $\eta/g_p$
($g_p$ is further coupled to $A$ via $\Phi_0=\ln (Ag_p)/\beta$, but this
can be considered as a consistency equation for $A$).

A typical model with $\beta\Psi (0)=8$, that gives a good fit to the surface
brightness structure of globular clusters, has $s_t=0.153$,
$\rho^*=4.677\times 10^4$, $M^*=0.1290$, $\sigma^{*2}=0.9650$. For
$M=6\times 10^5M_\odot$ and $\sigma^2=(7\mbox{\rm km/s})^2$ this gives
$\beta^{-1}=(7.126\mbox{\rm km/s})^2$, $\eta/g_p=2.10\times
10^{-4}M_\odot$pc$^{-3}$ (km/s)$^{-3}$, corresponding to a tidal radius
of 60.3~pc and a central density of $3.6\times 10^3 M_\odot$pc$^{-3}$.
Choosing $\eta$ to be higher than the maximal (central) phase space
density of the system 
($\eta >10M_\odot\mbox{\rm pc}^{-3}\mbox{\rm (km/s)}^{-3}$)
means that $g_p<5\times 10^4$. Thus statistical mechanics gives
reasonable results for a globular cluster of $6\times 10^5$ 1$M_\odot$
stars if one divides phase space in microcells smaller than $\approx
0.1 \mbox{\rm pc}^3\mbox{\rm (km/s)}^3$ 
and macrocells containing of order $10^4$ or less microcells.

\section{Discussion and conclusions}
It has been demonstrated in this investigation, that a
spherical stellar dynamical equilibrium configuration described by
Maxwell-Boltzmann statistics, contrary to the ``standard'' isothermal
sphere, has finite
mass. This is because phase space cells can be occupied by an integral
number of stars only, rather than the continuos distribution implicit in
the classical Maxwell-Boltzmann distribution. The complete
Simons-Maxwell-Boltzmann distribution (Simons 1994) explicitly takes
this into account, and in stellar dynamical applications this leads to
finite mass because one does not integrate fractionally occupied cells
over infinite space.

A new class of physically well-motivated models for globular clusters
were introduced based on the new distribution function. They resemble
King models because of some common features such as an energy cutoff and
a lowering of the classical Maxwellian, but in the King models
these features are not included in a quite correct manner (King
models still allow fractional occupation of phase space cells, and the
lowered isothermal sphere does not maximize entropy, which equation
(1) does for fixed gravitational potential, total energy and mass,
resembling the situation for globular clusters prior to core collapse).

The new models presented here fit the overall properties of 
globular clusters prior to core collapse well for
approximately the same choice of central potential as for King models.
They have constant ``temperature'' ($\beta^{-1}$), but in contrast to the
infinite mass isothermal sphere, this does not reflect itself in
constant velocity dispersion (cf.\ Figure 2). For large central
potentials the models get closer to the standard isothermal
sphere, and the velocity dispersion gets closer to
$\beta$. For very low central potential ($j_{max}=[\exp (\beta\Psi )]=1$) the
models go over into $n=3/2$ polytropes, but this is far below the range of
potentials giving reasonable fits to globular clusters.

Any use of distribution functions in stellar dynamics involves an
(often implicit) assumption about the ``graininess'' of phase space and
a choice of unit cell. Contrary to quantum mechanics, where Planck's
constant defines a fundamental cell size, there is no natural choice in
stellar dynamics. In fact there are even several ``schools'' arguing
either for a choice related to the observational resolution (interpreting
entropy as a measure of information content), or to the typical
phase space size of elements which are not disrupted by interactions
over the lifetime of the system studied.
The parameters extracted from fits to observations for the present model
include measures of the fundamental phase
space density, $\eta$, and the number of microcells in a macrocell,
$g_p$, but only in the combination $\eta /g_p$ (which can also be
interpreted as the mass of a fundamental particle (star) divided by the
volume of a macrocell). This could indicate, that there is a physically
motivated relation between the central potential (or more generally the
structure of the cluster) and the scale on which mixing in phase space
takes place. As discussed
above, values giving fits to globular cluster data seem reasonable, and
are not very dependent of the choice of model, as long
as $\beta\Psi (0)$ is not too small.
\bigskip

I thank Jens Hjorth for useful discussions. This work was supported by
the Theoretical Astrophysics Center under the Danish National Research
Foundation.
\newpage

\section*{References}

\begin{description}
\item Binney, J.,\& Tremaine, S.\ 1987, Galactic Dynamics
(Princeton: Princeton Univ.\ Press)
\item Elson, R., Hut, P.,\& Inagaki, S.\ 1987, ARA\&A, 25, 565
\item King, I.R.\ 1966, AJ, 71, 64
\item Mihalas, D.,\& Binney, J.J.\ 1981, Galactic Astronomy
(2d ed.; San Francisco: Freeman)
\item Simons, S.\ 1994, Am.\ J.\ Phys., 62, 515
\end{description}
\newpage

\section*{Figures}

{\bf Figure 1.} Density profiles of isothermal spheres with central
potential $\beta\Psi (0)=3$, 6, 9, 12 (solid curves), 
and $\infty$ (dotted curve;
equivalent to the classical, infinite mass isothermal sphere). For
comparison are shown the corresponding King models for the same choices
of $\beta\Psi (0)$ (dashed curves).

\bigskip\noindent
{\bf Figure 2.} One-dimensional root-mean-square velocity dispersion
profiles for the same models depicted in Figure 1. $v_{\rm
RMS}=\sqrt{\overline{v^2}(r)\beta/3}$.

\bigskip\noindent
{\bf Figure 3.} The concentration parameter $c\equiv\log_{10}(r_t/r_0)$
as a function of the dimensionless central potential for isothermal
spheres (solid curve) and King models (dashed curve).

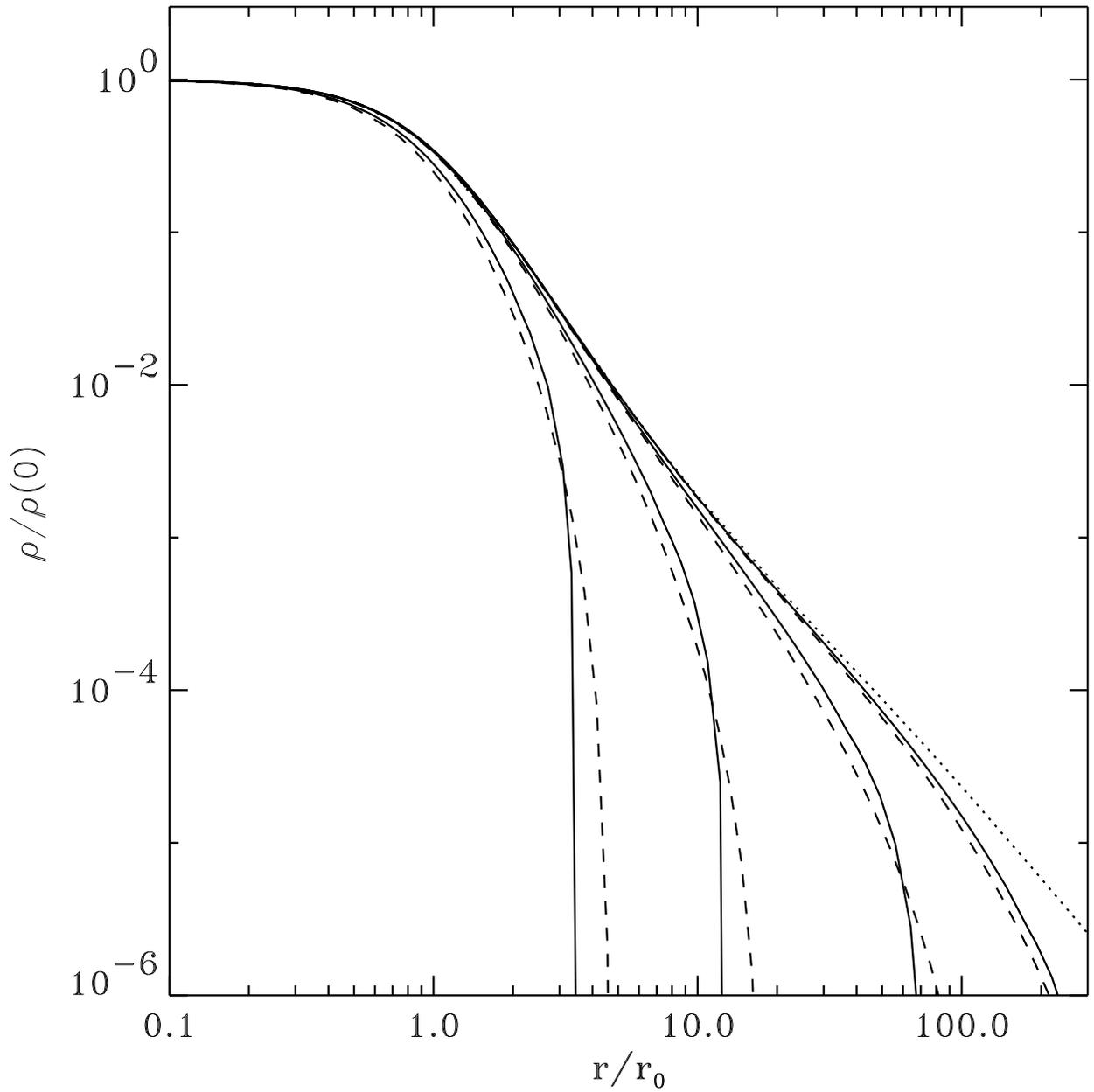
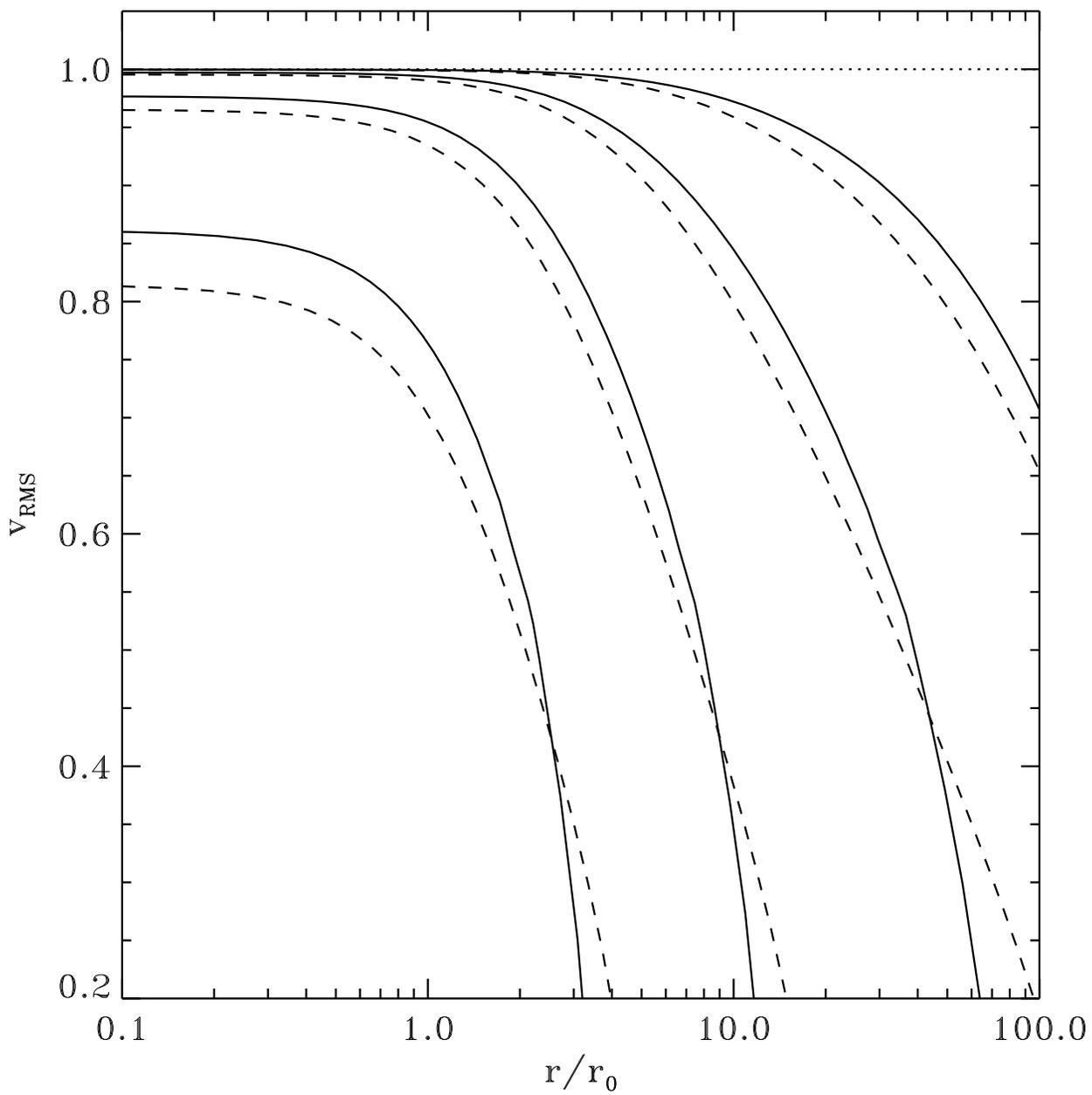
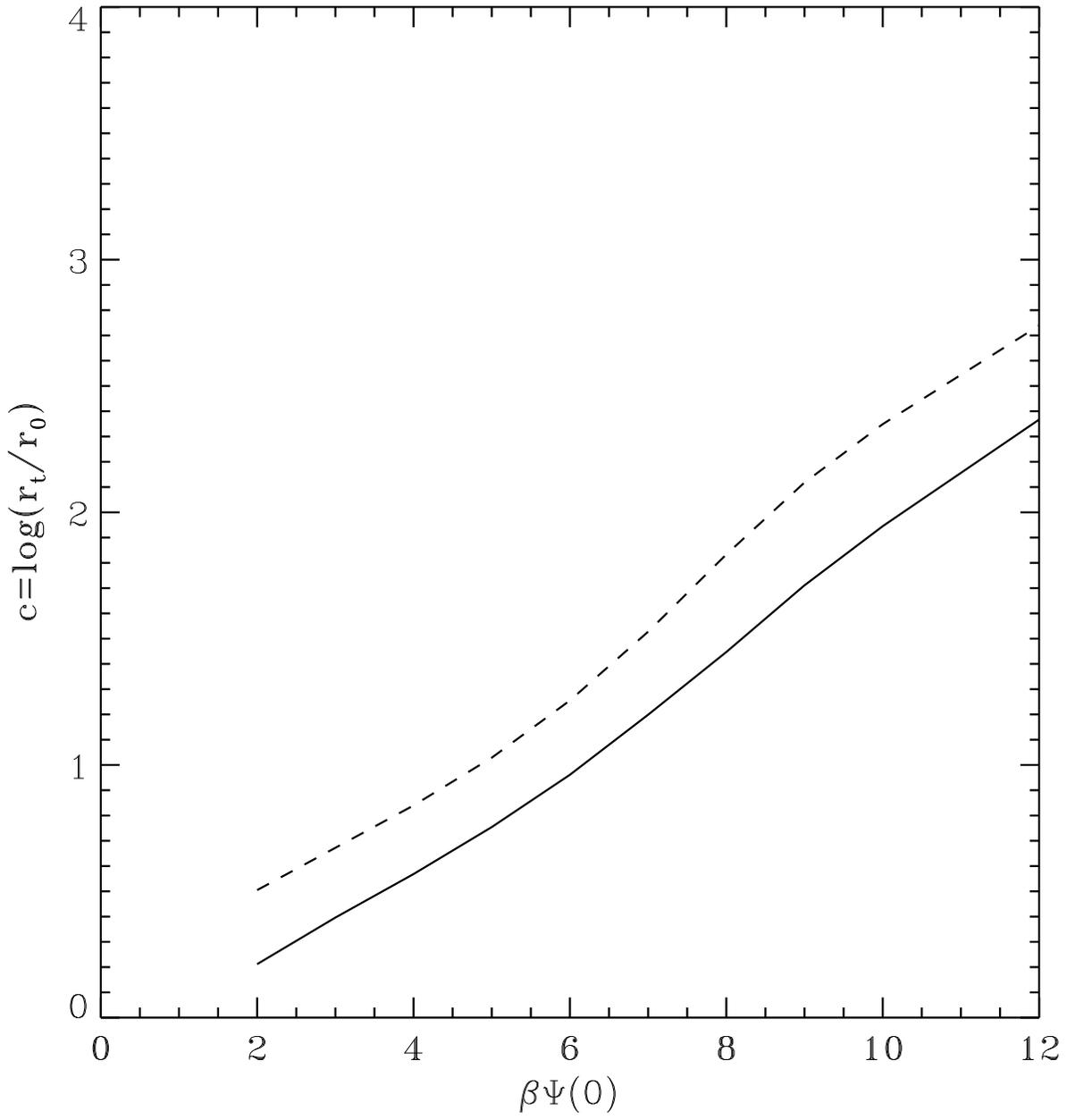
\end{document}